\documentclass[prl,amsmath,amssymb,showpacs,superscriptaddress,twocolumn]
{revtex4}
\usepackage{dcolumn,epsfig,color}
\begin{document} 

\title{Comment on \\ ``Test of the Brink-Axel Hypothesis for the Pygmy Dipole 
       Resonance'' \\[1mm]
$\left[{\rm D.~Martin}~{\it et~al.},~{\rm Phys.~Rev.~Lett.}~{\bf 119},~182503~(2017)\right]$}

\author{R.~Rusev} 
\affiliation{Los Alamos National Laboratory, Los Alamos, New Mexico 87545, USA}

\author{R.~Schwengner} 
\affiliation{Helmholtz-Zentrum Dresden-Rossendorf, 01328 Dresden, Germany}

\author{A.~R.~Junghans} 
\affiliation{Helmholtz-Zentrum Dresden-Rossendorf, 01328 Dresden, Germany}


\maketitle

In a recent letter \cite{mar17}, new data from a $(p,p')$ experiment for the
nuclide $^{96}$Mo are presented and a test of the Brink-Axel (BA) hypothesis in
the energy region of the pygmy dipole resonance (PDR) is performed by a
comparison of $\gamma$-ray strength functions and level densities deduced from
the excitation of nuclear states in this experiment with those obtained from
the deexcitation of states in a ($^3$He,$^3$He') experiment \cite{gut05,lar10}.
The good agreement of these quantities from excitation and from $\gamma$ decay
proves the validity of the BA hypothesis and shows that ($p,p'$) experiments
provide independent information about $\gamma$-ray strength functions and level
densities.

In addition, the new ($p,p'$) data are also compared with data obtained for
$^{96}$Mo by using bremsstrahlung in the ($\gamma,\gamma'$) reaction, also
called nuclear resonance fluorescence (NRF) \cite{rus09}. In this context,
the authors of Ref.~\cite{mar17} claim that there is an ``apparent violation of
the BA hypothesis in the low-energy regime suggested by the NRF data in
Ref.~\cite{rus09}''. This statement is not explained and is not correct. There
is no discussion of the BA hypothesis and its violation in Ref.~\cite{rus09}.
The simulations of statistical $\gamma$-ray cascades applied in the analysis of
the Mo isotopes \cite{rus08,rus09} and in following studies use the same input
strength functions for the photoexcitation of nuclear states and their
subsequent deexcitation. This means that they imply the validity of the BA
hypothesis, as described in Ref.~\cite{rus08}.

The strength function deduced from the $^{96}$Mo($\gamma,\gamma'$) data is
compared with that deduced from the ($p,p'$) data in Fig.~3 of
Ref.~\cite{mar17}. The authors of Ref.~\cite{mar17} notice that
``the ($\gamma,\gamma'$) data agree in the 7-8 MeV excitation energy region,
but clearly underestimate the present results at higher $E_x$.'' First, one
sees that also the ($\gamma,\gamma'$) data below 7 MeV agree with the ($p,p'$)
data within their uncertainties except for the value at about 6.7 MeV. The 
($\gamma,\gamma'$) data tend to be greater whereas the ($^3$He,$^3$He')
are smaller than the ($p,p'$) data. At $E_x$ higher than 8 MeV, the
($\gamma,\gamma'$) data are smaller than the ($p,p'$) data except for the value
at about 8.5 MeV while there are no ($^3$He,$^3$He') data. At the neutron
separation energy $S_n$ and above, the absorption cross section is the sum of
the cross sections of the ($\gamma,\gamma'$) and $(\gamma,n$) channels and,
consequently, this sum must be compared with the ($p,p'$) results. In the
present case of $^{96}$Mo, values for both the coexisting channels are
available at $E_x = S_n$. These are seen in Fig.~3(a) of Ref.~\cite{mar17} on
the vertical dashed-dotted line indicating $S_n$. The sum of these values
\cite{rus09,bei74} gives about $1.2 \times 10^{-7}$ MeV$^{-3}$, which agrees
well with the ($p,p'$) value. Considering all the facts just discussed, a
striking feature of the ($\gamma,\gamma'$) data that could be associated with a
violation of the BA hypothesis is not apparent from the comparison with the
($p,p'$) data.

We want to add that we performed combined studies using ($\gamma,\gamma'$) as
well as ($n,\gamma$) experiments for the final nuclei $^{78}$Se \cite{sch12},
$^{114}$Cd \cite{mas16}, and $^{196}$Pt \cite{mas13}. The respective targets
were chosen such that the ($n,\gamma$) capture states and the states populated
in ($\gamma,\gamma'$) both have spin $J$ = 1. A consistent description of the
spectra of the two reactions with identical strength functions for $\gamma$
absorption in ($\gamma,\gamma'$) and $\gamma$ decay in ($n,\gamma$) was
achieved, which is a clear confirmation of the validity of the BA hypothesis.

Summarizing, a violation of the BA hypothesis by $(\gamma,\gamma')$ data for
$^{96}$Mo \cite{rus09} as claimed but not justified by the authors of
Ref.~\cite{mar17} is in contradiction to the data analysis described in
Refs.~\cite{rus09,rus08} and cannot be concluded from that data. In addition,
the comparison of the $(\gamma,\gamma')$ data \cite{rus09} with the new
($p,p'$) data \cite{mar17} does not show significant discrepancies that may
serve as an argument for such a violation.


\begin{thebibliography}{99}

\bibitem{mar17}
D. Martin {\it et al.}, Phys. Rev. Lett. {\bf 119}, 182503 (2017).

\bibitem{gut05}
M. Guttormsen {\it et al.}, Phys. Rev. C {\bf 71}, 044307 (2005).

\bibitem{lar10}
A. C. Larsen and S. Goriely, Phys. Rev. C {\bf 82}, 014318 (2010). 

\bibitem{rus09}
G. Rusev {\it et al.}, Phys. Rev. C {\bf 79}, 061302 (2009). 

\bibitem{rus08}
G. Rusev {\it et al.}, Phys. Rev. C {\bf 77}, 064321 (2008).

\bibitem{bei74}
H. Beil, R. Berg\`{e}re, P. Carlos, A. Lepr\^{e}tre, A. De Miniac, and 
A. Veyssi\`{e}re,
Nucl. Phys. {\bf A227}, 427 (1974).

\bibitem{sch12}
G. Schramm {\it et al.}, Phys. Rev. C {\bf 85}, 014311 (2012).

\bibitem{mas16} 
R. Massarczyk {\it et al.}, Phys. Rev. C {\bf 93}, 014301 (2016).

\bibitem{mas13} 
R. Massarczyk {\it et al.}, Phys. Rev. C {\bf 87}, 044306 (2013).


\end{thebibliography}
\end{document}